\documentclass[10pt,a4paper]{article}
\usepackage{amsmath}
\usepackage{amsthm}
\usepackage{amsfonts}
\usepackage{multicol}
\usepackage[dvips]{graphicx}
\usepackage{fixltx2e}

\newcommand{\danger}[1]{\textbf{#1}}

\input{epsf}             

\begin{document}

\title{\danger{The four particles paradox in special relativity}}
\author{\centerline{\danger{J. Manuel Garc\'\i a-Islas \footnote{
e-mail: jmgislas@iimas.unam.mx}}}  \\
Departamento de F\'\i sica Matem\'atica \\
Instituto de Investigaciones en Matem\'aticas Aplicadas y en Sistemas \\ 
Universidad Nacional Aut\'onoma de M\'exico, UNAM \\
A. Postal 20-726, 01000, M\'exico DF, M\'exico\\}

\maketitle

\begin{abstract}
We present a novel paradox in special relativity together with its solution. We call
it the four particles paradox. The purpose of this paradox is pedagogical and therefore 
directed towards students and lecturers
of physics. Even if most paradoxes in special relativity theory are very interrelated and
some are special cases of others, the paradox we present here is original and
illuminates on the very nice subject and the literature of special relativity. 
\end{abstract}

\section{Introduction}

Ever since its appearance $\cite{e}$, Einstein's special relativity theory has been filled with interesting paradoxes.
We couldn't agree more with Bernard Schutz's \cite{bs} when in his opinion paradoxes do not exist, as
these are only misunderstood problems.\footnote{See reference \cite{bs}, pages 23-24}

There may only be two reasons about the existence of many paradoxes of special relativity in the literature. 
These are only
misunderstood problems from a superficial knowledge of the subject, or they are posed by lecturers and researchers
in depth knowledge of the subject who
are interested in illustrating these problems to students of physics, like in \cite{r}, 
\cite{gj}, \cite{g}, \cite{cg}.

From this latter perspective, we can say that paradoxes in special relativity are interesting
problems which are at first confusing, wrongly pointing to inconsistencies with the theory, 
but that after a better understanding of the subject, they are finally very good exercises for students to master
the subject. 
 
In this work, we present a novel paradox along with its solution. We call it the four particles paradox.

The main purpose of this work is
at the pedagogical level, and will be very useful and a very nice example for students as well as 
for lecturers in relativity theory. Moreover, the paradox along with its solution 
requires elementary concepts of special relativity only.

\section{The paradox}

We now present the paradox, and its solution.
We invite the student to think about it before reading the solution. 

\bigskip

We will consider inertial frames which we denote $S$, $S'$ and $S''$. 
Mathematically, let us consider that points at inertial frames are given coordinates 
$(x,y,z)$, $(x' , y', z')$ and $(x'' , y'' , z'')$ respectively. We also suppose that they all move
with respect to each other along the $x, x' , x''$ direction and that all their axes are parallel.\footnote{
It is important to mention that we only need two spatial dimensions to describe the problem. However, we stick
to three spatial dimensions for aesthetic reasons. Just 
because physical objects such as trains, spaceships, cars, which are represented
by inertial frames, are three dimensional.}

\bigskip

Let us pose the 'paradox' 

\bigskip

\danger{The four particles paradox:} Two inertial frames $S$ and $S'$ move towards each other with respect to
an inertial frame $S''$ and with the same speed $v$ as measured by
$S''$. Eva (an observer) at rest in $S$ places two classical particles in her frame, one located at 
$A=(x_1,y_1,z_1)=(0,0,0)$ and the other at $B=(x_2,y_2,z_2)=(\ell, 0 , d)$. Manuel (an observer) at rest in $S'$ 
places two identical particles to Eva's in his frame, one located at 
$A'=(x_1 ',y_1 ',z_1 ')=(0' ,0' , 0')$ and the other at $B'=(x_2 ',y_2 ',z_2 ')=(\ell', 0' , d')$, 
such that $\mid \ell \mid = \mid \ell' \mid$ 
and $\mid d \mid = \mid d' \mid$. (See Figure 1).

\bigskip

The experiment consists of the following: 

\bigskip 

According to Eva the identical particles $B$ and $B'$ will collide and vanish\footnote{Throughout this article, particles 
will refer to classical particles, not to quantum ones. And when we say that they vanish as they collide, it means that
they will scatter and the observer will no longer see them.}
earlier than the identical particles 
$A$ and $A'$ because of length contraction along $x, x'$. (See Figure 2).
However, just after the collision of particle $B$ and 
$B'$, she decides to collect particle $A$
before it collides with particle $A'$. 

Analogously, to Manuel the identical particles $A$ and $A'$ are the ones which will collide and vanish 
earlier than the identical particles 
$B$ and $B'$, because of length contraction along $x, x'$. (See Figure 3). 
However, just after 
the collision of particle $A$ and $A'$, he decides to collect particle $B'$
before it collides with particle $B$.

To an anonymous observer at $S''$ the four particles will collide and vanish simultaneously and neither Eva
nor Manuel will have their corresponding particles in their hands.

\begin{figure}[h]
\begin{center}
\includegraphics[width=1.2\textwidth]{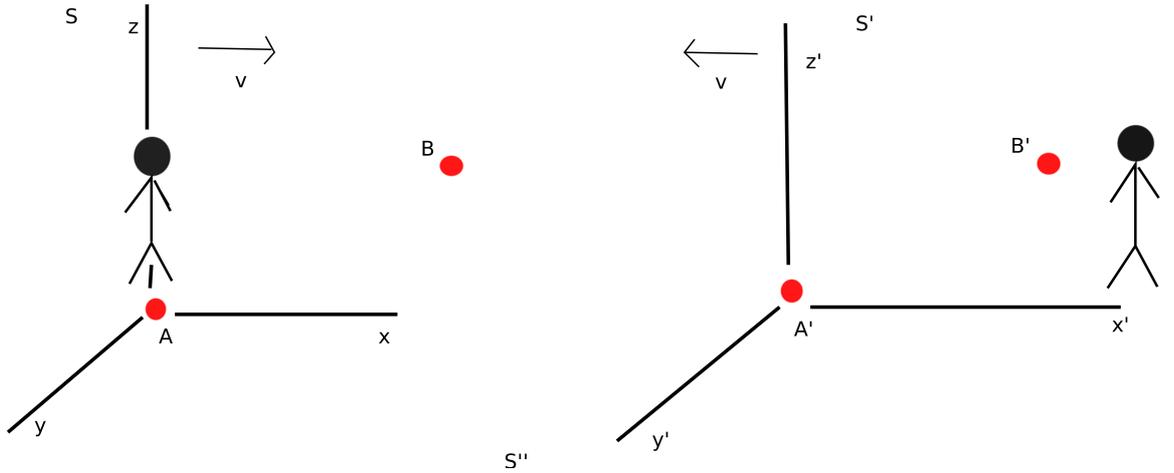}
\caption{Inertial systems $S$ and $S'$ moving towards each other at speed $v$, as seen from an inertial frame
$S''$. Particles $A$ and $B$ are drawn as seen by observer at $S$ and particles $A'$ and $B'$ are drawn as
seen by observer at $S'$.}
\end{center}
\end{figure}

So, how is it possible that Eva has in her hand particle $A$ if Manuel saw it vanished when it hit particle $A'$ ?
In the same way, how is it possible that Manuel has in his hand particle $B'$ if Eva saw it vanished when it hit particle $B$?
How is it possible that to the anonymous observer neither Eva nor Manuel have a particle in their hands.

Who is right? In other words;
Eva will claim she has the $A$ particle in her hand and that particle $B$ and $B'$ have vanished.
Manuel will claim he has the $B'$ particle in his hand and that particle $A'$ and $A$ have vanished.
The anonymous observer will claim the four particles have vanished.

\subsection{The solution}

Let us now present the solution.\footnote{We insist one more time to the student to think of the solution before
reading it.} We will use basic special relativity concepts only.

Due to the addition of velocities in special relativity,
Eva and Manuel are moving towards each other at speed

\begin{equation}
w= \frac{2v}{1+v^2} 
\end{equation}
According to Eva, particles at Manuel's frame are longitudinally separated a distance

\begin{equation}
L= \ell' \sqrt{1-w^2} 
\end{equation}
due to length contraction along the direction of motion. Moreover, they are vertically separated a distance
$\mid d \mid = \mid d' \mid$, since there is no contraction along the perpendicular direction of motion. 

Therefore, according to Eva, the identical particles $B$ and $B'$ will collide and vanish earlier than the identical particles 
$A$ and $A'$. 
Just after the collision of particle $B$ and 
$B'$, she decides to collect particle $A$
before it collides with particle $A'$. 

\begin{figure}[h]
\begin{center}
\includegraphics[width=1.2\textwidth]{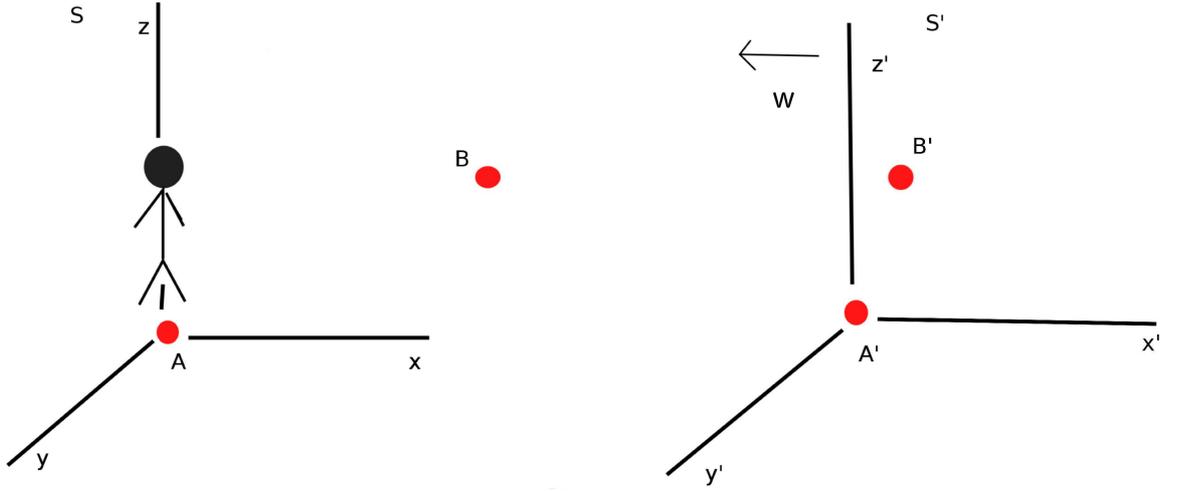}
\caption{Inertial system $S'$ moving towards $S$ at speed $w$. Particles $A'$ and $B'$ as
seen by observer at $S$.}
\end{center}
\end{figure}

Analogously, Manuel will observe particles at Eva's frame longitudinally separated a distance

 \begin{equation}
L'= \ell \sqrt{1-w^2} 
\end{equation}
due to length contraction along the direction of motion. Moreover, they are vertically separated a distance
$\mid d \mid = \mid d' \mid$, since there is no contraction along the perpendicular direction of motion. 

Therefore, according to Manuel, the identical particles $A$ and $A'$ will collide and vanish earlier than the identical particles 
$B$ and $B'$. 
Just after the collision of particle $A$ and 
$A'$, he decides to collect particle $B'$
before it collides with particle $B$. 

\begin{figure}[h]
\begin{center}
\includegraphics[width=1.2\textwidth]{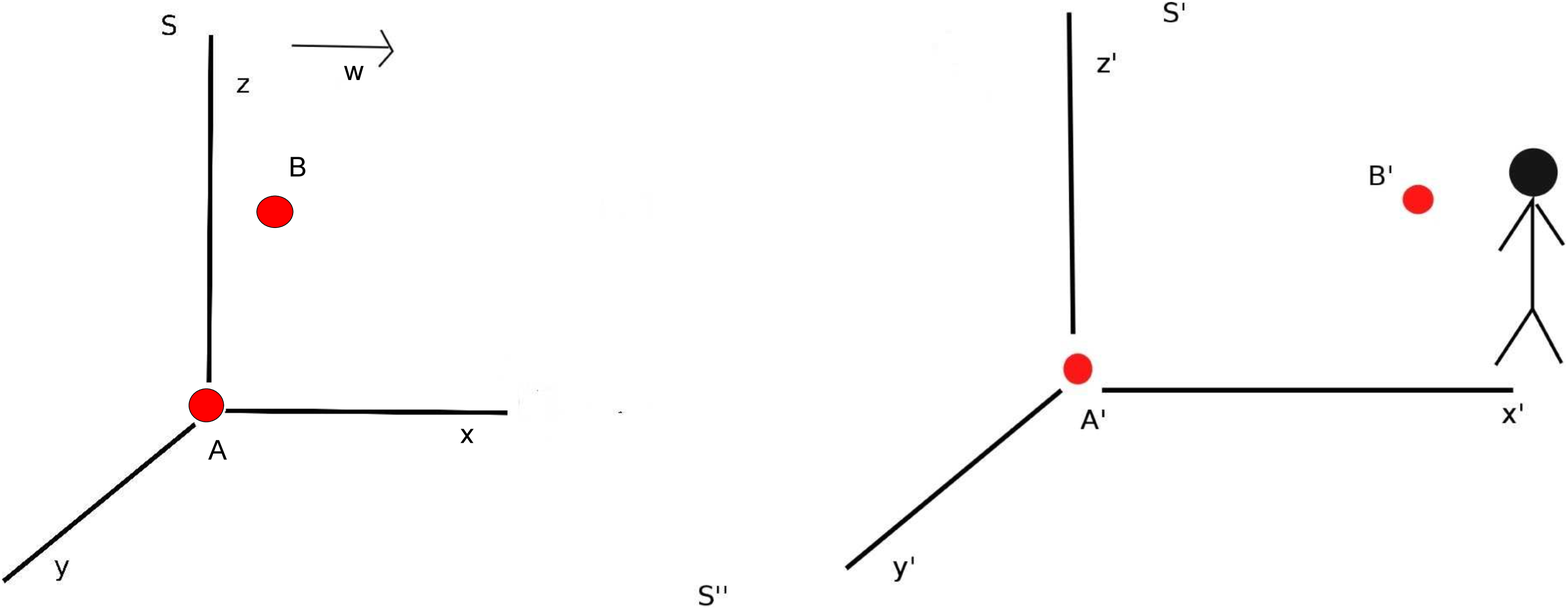}
\caption{Inertial system $S$ moving towards $S'$ at speed $w$. Particles $A$ and $B$ as
seen by observer at $S'$.}
\end{center}
\end{figure}

\bigskip

Let us now see that it is not possible that Eva collects particle $A$ before it collides with particle $A'$. 
Particles $A$ and $A'$ will collide and vanish before she prevents them from colliding.
And the same applies to Manuel, it is not possible that he collects particle $B'$ before it collides with particle $B$. 
Particles $B$ and $B'$ will collide and vanish before he prevents them from colliding.

\bigskip

If Eva were located just where her $A$ particle is situated\footnote{Like sitting on top of it, so that she collects it as fast
as possible.}, then this is what happens. Recall that in special relativity all signal information
is transmitted, at most, at the speed of light. 
Therefore, when particle $B$ and $B'$ collide and vanish, a clock situated at the point of collision
will read $t_0= 0$. Then, Eva will have knowledge
of this collision when light coming from the point of collision gets to her.

The point of collision of particles $B$ and $B'$ is separated from particle $A$ a distance $r=\sqrt{\ell^2 + d^2}$.
Therefore,
information about the collision of particles $B$ and $B'$
will reach Eva\footnote{In units where where $c=1$} at proper time $t_1 = \sqrt{\ell^2 + d^2}$. It will be enough
to consider the longitudinally separation of the point of collision of particles $B$ and $B'$ and particle $A$
given by $\ell$ so that information of the collision of particles $B$ and $B'$ will reach Eva
at proper time $t_1= \ell < \sqrt{\ell^2 + d^2}$.

From Eva's point of view, at the moment of collision of particles $B$ and $B'$, particles $A$ and $A'$ are 
longitudinally located
a distance $D$ apart given by

\begin{equation}
D= \mid \ell \mid - \mid L \mid = \mid \ell \mid - \mid \ell' \mid \sqrt{1-w^2} 
= \mid \ell \mid - \mid \ell \mid \sqrt{1-w^2} = \mid \ell \mid [1- \sqrt{1-w^2}] 
\end{equation}
and therefore particles $A$ and $A'$ will collide and vanish at Eva's proper time given by

\begin{equation}
t_2= \frac{D}{w} = \frac{\ell [1- \sqrt{1-w^2}]}{w} 
\end{equation}
It can easily be checked that $t_2 < t_1$. Let us check this strict inequality

\begin{eqnarray}
\frac{\ell [1- \sqrt{1-w^2}]}{w} < \ell  \nonumber \\
\Rightarrow \ \ \  [1- \sqrt{1-w^2}] < w \   \nonumber \\
\Rightarrow \ \ \ - \sqrt{1-w^2} <  w-1 \nonumber \\
\Rightarrow \ \ \ \ \ 1-w^2 > [1-w]^2 \nonumber \\
\Rightarrow \ \ \ \ \ \ \ \ \ \ 0 > 2w [w -1] 
\end{eqnarray}
and this latter inequality is true, since $w<1$. 

\bigskip

Therefore, particles $A$ and $A'$ will collide and vanish before Eva knows
that particles $B$ and $B'$ have collided, and therefore, she cannot collect particle
$A$ before it collides with particle $A'$. By the time she knows that particle
$B$ and $B'$ have collided, particles $A$ and $A'$ will also be vanished.  

The same method applies to Manuel with the conclusion that he will not be able
to collect particle $B'$ before it collides with particle $B$, since by the time he realises about
the collision of particles $A$ and $A'$, particles $B$ and $B'$ will be vanished.

The paradox is solved. Neither Eva, nor Manuel will have collected a particle, thus agreeing with the
anonymous observer.

\bigskip

The paradox we presented here can be seen as 
a smart variation of the two colliding inclined rods paradox\footnote{Compare with \cite{lp}.} presented
in $\cite{lp}$. 
However the solution presented here deals with pure simple relativistic concepts.
It does not involve the idea of 
'extended present' as invoked to solve the paradox in $\cite{lp}$. 
In our opinion the term 'extended present' does not exist. 
The solution we presented here solves both, 
the four particles paradox 
and the one presented in \cite{lp}.  

It can easily be seen that in terms of the space-time geometry  
the observer at $S$ concludes that the separation of the events corresponding to the collision of particles
$B$ and $B'$ and the collision of particles $A$ and $A'$ is space-like, as well as the observer
at $S'$ concludes that the separation of the events corresponding to the collision of particles
$A$ and $A'$ and the collision of particles $B$ and $B'$ is also space-like.

It is a trivial exercise (for students) to find the Lorentz transformation between 
the inertial frame $S$ and $S'$ which sends the space-like separated events 
at $S$ into the space-like separated events at $S'$.
Recall that Lorentz transformations send space-like vectors into 
space-like vectors.\footnote{It also sends time-like vectors into time-like vectors and null vectors into null vectors.}    

To sum up, the paradox has been solved using only basic concepts of special relativity, and it is suitable
to be presented as a good exercise for students. It illuminates on the subject of relativity and can be used 
at the pedagogical level by teachers in the area.

\end{document}